\documentstyle[prl,aps,epsfig]{revtex}
\begin{document}\hbadness=10000
\title{Strangeness Enhancement and Canonical Suppression}
\author{Ahmed Tounsi$^{\rm (a)}$, Krzysztof Redlich $^{\rm (b,c)}$} 
\address{
$^{\rm (a)}$ Laboratoire de Physique Th\'eorique et Hautes Energies,
Universit\'e Paris 7, 2 pl Jussieu, F--75251 Cedex 05\\
$^{\rm (b)}$ Theory Division, CERN, CH-1211 Geneva 23,
Switzerland\\
$^{\rm (c)}$ Institute of Theoretical Physics, University of Wroclaw,
PL-50204 Wroclaw
}
%
\maketitle
\begin{abstract}
We demonstrate the essential role of canonical suppression in strangeness
enhancement. The  pattern of enhancement of strange and multistrange
baryons observed by the WA97 collaboration can be understood
on this basis. Besides, it is shown  that in canonical approach strangeness enhancement is a decreasing function of collision energy. It is the largest at
$\sqrt{s}=8.7$ GeV where
the enhancement of $\Omega,\Xi$ and $\Lambda$  is of the order of
100, 20 and 3 respectively.
\end{abstract}
\pacs{PACS: xxx, yyyy}
\section{introduction}
Ultra-relativistic heavy ion collisions provide a unique
opportunity to study the properties of  highly excited hadronic
matter under extreme conditions of high density and temperature
\cite{qm,satz,stoc,stachel,heinz1}. From the analysis of rapidity
distribution of protons and of their transverse energy measured in
158 A GeV/c Pb+Pb collisions an estimate of the initial condition
 leads to energy density of 2-3
GeV/fm$^3$ and a baryon density  of the order of 0.7/fm$^3$.
Lattice QCD at vanishing baryon density suggests that the phase
transition from confined to the quark-gluon plasma (QGP) phase
appears at the temperature $T_c=173\pm 8$ MeV which corresponds to
the critical energy density \cite{karsch} $\epsilon_c\sim 0.6\pm 0.3$ GeV/fm$^3$. One could thus conclude that the required initial
conditions for quark deconfinement are already reached in heavy
ion collisions at top SPS energy  and RHIC energies. Thus, of particular
relevance was to find experimental probes to check whether the
produced medium in its early stage was indeed in the QGP phase.
Different probes have been theoretically proposed and studied
in relation with CERN-SPS and more recently with BNL-RHIC experiments.
We will concentrate on strange hadrons and particularly on 
multistrange baryons. More precisely we will examine the issue
of enhancement of strange and multistrange baryons  as a possible signature of QGP formation.

\section{Strangeness enhancement and QGP}
It was long ago argued that enhanced production of strange
particles in nucleus-nucleus (AA) collisions, relatively to proton-proton (pp)
or to proton-nucleus (pA) collisions, could be a signal of QGP 
formation \cite{rafelski1}. In QGP the production and
equilibration of strangeness is very efficient due to  a large
gluon density and a low energy threshold for dominant QCD
processes of  $s\bar s$ production \cite{rafelski1,raf}:
\begin{equation}
GG \to s\bar s\nonumber
\end{equation}
 In contrast, in a 
hadronic system, e.g., in pp, the
higher threshold for strangeness production  was argued
to make the strangeness yield considerably
smaller and the equilibration time much longer than in QGP. From these strangeness QGP  characteristics one expects
a {\it global strangeness enhancement\/},
 which should increase from pp, pA to AA collisions, as well as
 {\it enhancement of multistrange (anti)baryons\/}. The global strangeness
content in the collision fireball  is measured by
 $<s\bar s>/N_{part}$, the total number of strange quarks per
 participant nucleon  or by
$ <s\bar s>/<u\bar u +d\bar d>$, the total number of strange
quarks per produced light quark. Furthermore,
this  (anti)hyperon enhancement was predicted to depend on the strangeness content
of the (anti)hyperons and to appear in a typical hierarchy:
$$E_{\Lambda}  < E_{\Xi} < E_{\Omega}$$
  This
hierarchy is observed by the WA97 and NA57  collaborations \cite{andersen,carrer} which measure
 the yield per participant in Pb+Pb relative to p+Pb and p+Be collisions. In particular the WA97 data show a pattern with
this hierarchy and a saturation of enhancement for a number
of participant nucleons $N_{part} > 100$.  Recent results   of the NA57 collaboration\cite{carrer} are
showing in addition an abrupt  change of anti-cascade enhancement
for lower $N_{part}$. Similar behavior  was previously seen on the
 $K^+$ yield measured by the NA52 experiment in Pb-Pb
collisions \cite{kabana}. These results are very interesting as
they might be interpreted as an indication of the onset of new
dynamics. However, a more detailed experimental study and
theoretical understanding are still required here. It is, e.g., not
clear  what is the origin of different centrality dependence of
$\Xi$ and $\bar\Xi$. Nevertheless, this abrupt change could
be possibly accounted for in canonical approach 
by assuming a
very particular increase of volume parameter with centrality
\cite{muller}.

Anyway, strangeness enhancement is also seen at low energies,
 and found to be a decreasing function of collision energy
in  a compilation of the data on $K^+/\pi^+$ ratio
in A+A relative to p+p collisions, where the double ratio
$(K^+/\pi^+)_{(AA)}/(K^+/\pi^+)_{pp}$ can be considered as
an enhancement measure \cite{ogilvie,blume}. Such
a behaviour with collision energy could also be expected
for multistrange baryons. It was indeed shown 
\cite{hamieh,red2} that a statistical model (SM) implementing
canonical strangeness conservation
explains the WA97 pattern and predicts that
enhancement is a decreasing function of collision
energy. This, we summarize in the following sections.

\section{Strangeness enhancement and canonical suppression}
 The enhancement $E$ measured in experiments is the ratio of the yield
of a given (anti)baryon per participant nucleon in the large AA system
to the yield of the same (anti)baryon in a the small pp or pA
system:
\begin{equation}
 E=\frac{\left(Yield\right)_{\vert AA}}{\left(Yield\right)_{\vert pA}}\label{E}
\end{equation}
In a large system with a large number of produced particles,
the conservation law of a quantum number, e.g., strangeness,
can be implemented on the average by using the corresponding
chemical potential. This is the Grand Canonical formulation (GC). In a small system, however, with small
particles multiplicities, conservation laws must be implemented
locally on an event-by-event basis\cite{hagedorn,ko}.
This is the Canonical formulation (C). The (C) conservation
of quantum numbers is known to severely reduce the phase space
available for particle production \cite{hagedorn}. This the canonical suppression (CS). If in Eq.\,(\ref{E}) the denominator is reduced by CS then
E is increased. That is, in our approach, the essence of strangeness
enhancement from pp, pA to AA collisions. 

To better understand  CS, consider a gas of particles with
strangeness $s=+1, 0,-1$ and with  {\it total\/} strangeness $S=0$
(this condition is imposed by the fact that in heavy ion collisions
the initial state of the system is strangeness neutral).
Group theory projection techniques \cite{red3,tou} allow
to construct the partition function of the gas, from which
all thermal physical quantities are derived.
One obtains for the thermal kaon density in the (C) formulation  
\begin{equation}
n_K^C=\frac{Z_K^1}{V}\frac{S_1}{\sqrt{S_1S_{-1}}}\frac{I_1(x_1)}{I_0(x_1)}
\label{kaon}
\end{equation}
where
\begin{equation}
 Z_K^1= V{{g_K}\over {2\pi^2}}
m_k^2TK_2\left({{m_K}\over T}\right)\label{z1}
\end{equation}
\begin{equation}
S_1=Z^1_K+Z^1_{\bar\Lambda}+Z^1_{K^{\star}}+...\label{s1}
\end{equation}
\begin{equation}
x_1\equiv 2\sqrt{S_1S_{-1}}\propto V\label{x_1}
\end{equation}
$V$ is the volume  parameter which is assumed, as usual, to be linear in
$N_{part}$:
\begin{equation} 
 V =\frac{V_0}{2} N_{part}\label{vol}
\end{equation}
where $ V_0\approx 7\,{\rm fm}^3$ is taken as the volume of the nucleon.
$S_1$ and $S_{-1}$ are the sum over one-particle partition functions for
particles carrying strangenes 1 and -1 respectively.
 $ I_1,I_0$  and $ K_2$
are modified Bessel and Hankel functions.
The density $n_K^C$ in Eq.\,(\ref{kaon}) depends on the volume, but
for $x_1 \to \infty$, $I_1(x_1)/I_0(x_1)\to 1$ and
$n_K^C$ reaches its GC limit, $ n_K^{GC}$, independent of the volume.
For $x_1 \to 0$  $I_1(x_1)/I_0(x_1)\to x_1/2$ and
the density $n_K^C$ is linearly dependent on the volume.
One clearly sees on Eq.\,(\ref{kaon}) that the factor
\begin{equation}
  F_{CS1}=\frac{I_1(x_1)}{I_0(x_1)}\label{fcs}
\end{equation}
called canonical suppression factor,
measures the deviation of the kaon  density from its GC
value $n_K^{GC}$.
\begin{figure}[htb]
\vspace*{-1cm}
\begin{minipage}[t]{75mm}
\epsfig{width=7.5cm,figure=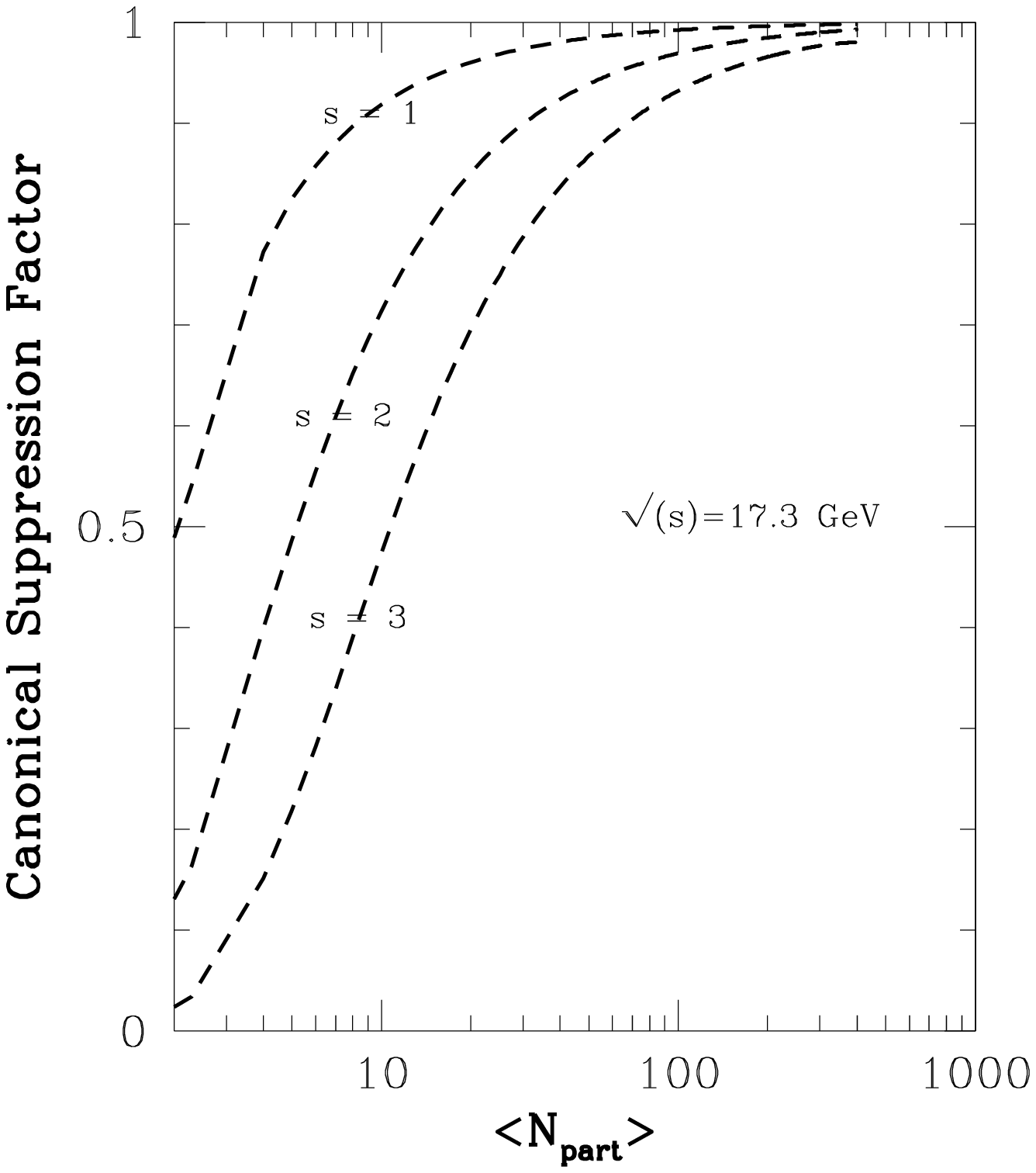}
\vskip -0.4cm
\caption{Canonical suppression factor for three values of particle strangeness:
$s=1,2,3$, at top CERN-SPS energy. \protect\label{reduc123}}
\end{minipage}
\hspace{\fill}
\begin{minipage}[t]{76mm}
\epsfig{width=7.5cm,figure=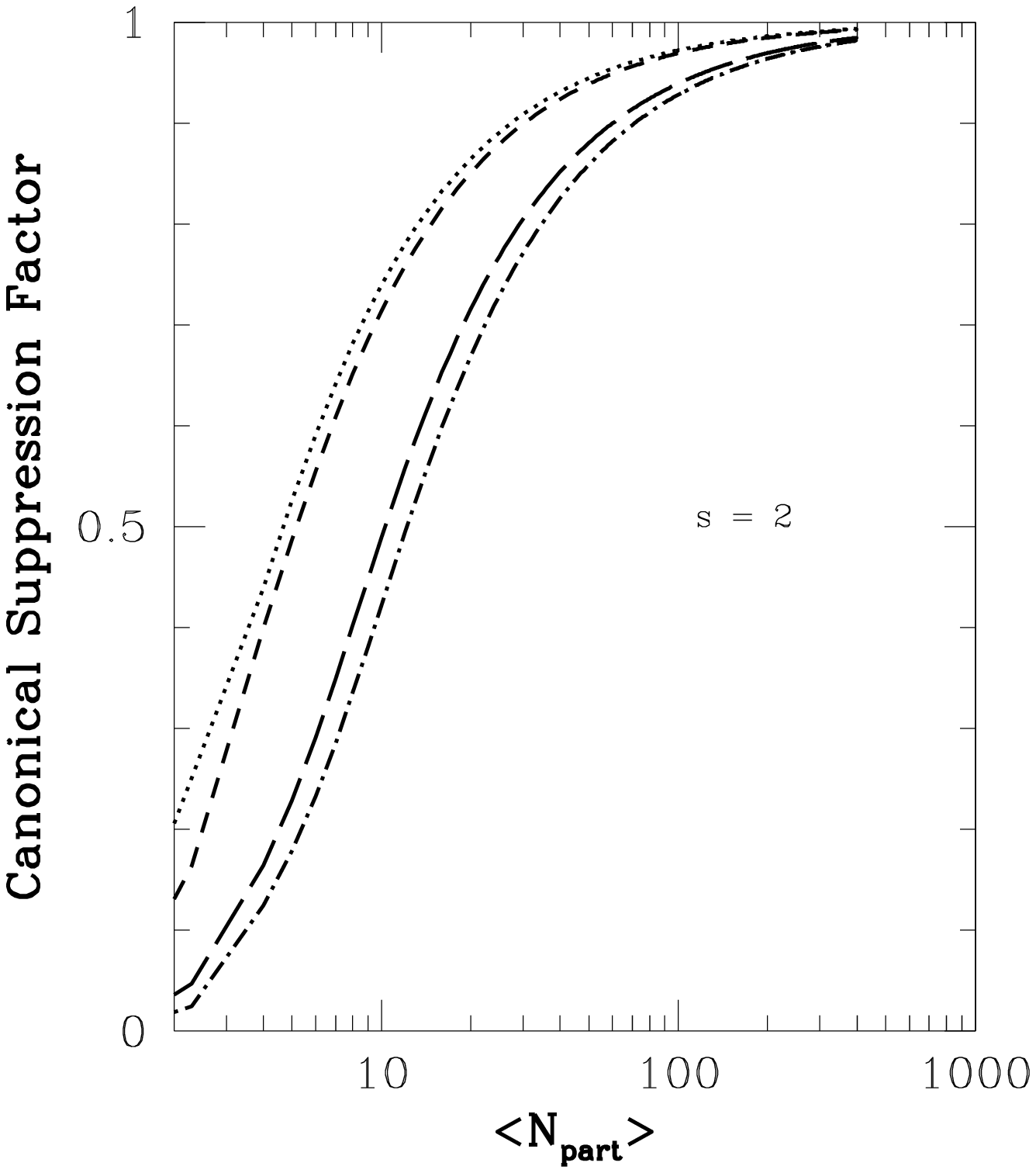}
\vskip -0.4cm
\caption{ 
dotted line: $\sqrt{s}=130$GeV; short dashed line: $\sqrt{s}= 17.3$GeV; 
long dashed line: $\sqrt{s}=12.3$GeV; dot-short dashed line: $\sqrt{s}=8.3$GeV
 \protect\label{reduc2}}
\end{minipage}
\end{figure}

The previous considerations can naturally be extended to 
a gas of particles with strangeness $s=0,\pm 1,\pm 2, \pm 3$, 
i.e., to a gas composed of all particles (antiparticles) and their
resonances.
 With the condition
that the gas has total strangeness $S=0$,
the canonical particle density of a particle $i$ having strangeness
$s$ reads \cite{newpbm}
\begin{equation}
n_{i}^C=\frac{1}{V}{{Z^1_{i}}\over {Z_{S=0}^C}}
\sum_{n=-\infty}^{\infty}\sum_{p=-\infty}^{\infty} a_{3}^{p}
a_{2}^{n}
 a_{1}^{{-2n-3p- s}} I_n(x_2) I_p(x_3) I_{-2n-3p- s}(x_1)$$
  \label{dens}
\end{equation}
where
\begin{eqnarray}
a_i&=& \sqrt{{S_i}/{S_{-i}}}\label{ssi}\\
x_i &=& 2\sqrt{S_iS_{-i}}\propto V\label{xii}\\
Z^C_{S=0}&=& \sum_{n=-\infty}^{\infty}\sum_{p=-\infty}^{\infty} a_{3}^{p}
a_{2}^{n} a_{1}^{{-2n-3p}} I_n(x_2) I_p(x_3) I_{-2n-3p}(x_1)
  \label{partf}
\end{eqnarray}
 In Eq.\,(\ref{ssi}),   $S_i =\sum_k Z^1_k $ where the sum is
over all particles and resonances carrying strangeness $i$.
For a particle of mass $m_k$, with spin-isospin degeneracy
factor $g_k$, carrying baryon number $B_k$ and electric charge $Q_K$,
with corresponding chemical potential $\mu_B$ and $\mu_Q$, the
one-particle partition function is given, in the Boltzmann,  by
\begin{equation}
Z_k^1\equiv V{{g_k}\over {2\pi^2}}
m_k^2TK_2\left({{m_k}\over T}\right)\exp (B_k\mu_B+Q_k\mu_Q)
\label{zzk}
\end{equation}
Here too one can show \cite{hamieh} that for small $x_1$
\begin{equation}  
 n_{i}^C\simeq\frac
{Z^1_{i}}{V} {{(S_{1})^s}\over 
{{(S_{+1}}S_ {-1})^{s/2}}}~
      { { I_s(x_1)}\over {I_0(x_1)}}
\end{equation}
and the canonical suppression factor is now
\begin{equation}
 F_{CSs}=\frac{ I_s(x_1)} {I_0(x_1)}\label{fcs1}
\end{equation} 
In Eq.\,(\ref{fcs}) and Eq.\,(\ref{fcs1}) one 
sees that  strangeness content of the particle appears
in  the suppression factor as the order of Bessel function
$I_s(x_1)$. Fig.\,\,\ref{reduc123} shows
that the suppression factor, for a given value of $N_{part}$,
is smaller for larger values of $s$. At a given energy (temperature)
$N_{part}$  depends on the colliding system. In the small system for
p-p collisions $N_{part}=2$. For p-Be (p-Pb) collisions 
$N_{part}\approx 2.5\,\,(\approx 4.75)$. These values have been determined
by the WA97 collaboration from a Wounded Nucleon Model in the
framework of the Glauber Model\cite{carrer1}.    
 In particular for small
$x_1$, $F_{CSs} \sim (x_1/2)^s$, and one expects that the larger  the strangeness content of
the particle the smaller the suppression factor and hence
the larger the enhancement. This explains\cite{hamieh}
the enhancement hierarchy of the WA97 pattern. Furthermore,
 one sees on Fig.\,\,\ref{reduc2} that for a given strangeness, e.g., $s=2$,
the suppression factor, for any number of participant nucleons, is decreasing
with decreasing  energy: this means that  enhancement increases with decreasing  energy.  
Finally, enhancement saturation appears, as explained above,
as the grand canonical limit for large number of participant nucleons (large volume).

\section{Strangeness enhancement energy dependence}
We have studied strangeness enhancement at four energies:
$\sqrt s = 8.73, \ 12.3 , \ 17.3$ GeV  (NA49 and WA97, SPS) and
$\sqrt s = 130$ GeV (RHIC). To study the energy and centrality dependence of (multi)strange particle yields in terms of the above model one  needs to establish first the variation  of  thermal parameters with energy
and centrality. Temperature is to a good approximation\cite{b3} only a function of collision energy and is independent of the number of
participating nucleons. Baryonic chemical potential $\mu_B$ is
weakly dependent on centrality\cite{red2}. Thermal parameters are, however
sensitive to collision energy. At top SPS energy $\sqrt{s}=17.3$ GeV
we use the parameters obtained \cite{braun} in experimental data
analysis: $T=166$ MeV and  $\mu_B=266$ MeV. At RHIC energy  we take \cite{braun1} $T= 175$ MeV and $\mu_B=51$ MeV. At $\sqrt s = 8.73, \ 12.3$ GeV
the parameters are fixed according to the method explained in 
\cite{red2}. We have $T=145$ MeV, $\mu_B=370$ MeV and 
$T=152$ MeV , $\mu_B=280$ MeV respectively.

\begin{figure}[htb]
\vskip -1.cm
\begin{minipage}[t]{75mm}
\epsfig{width=7.4cm,figure=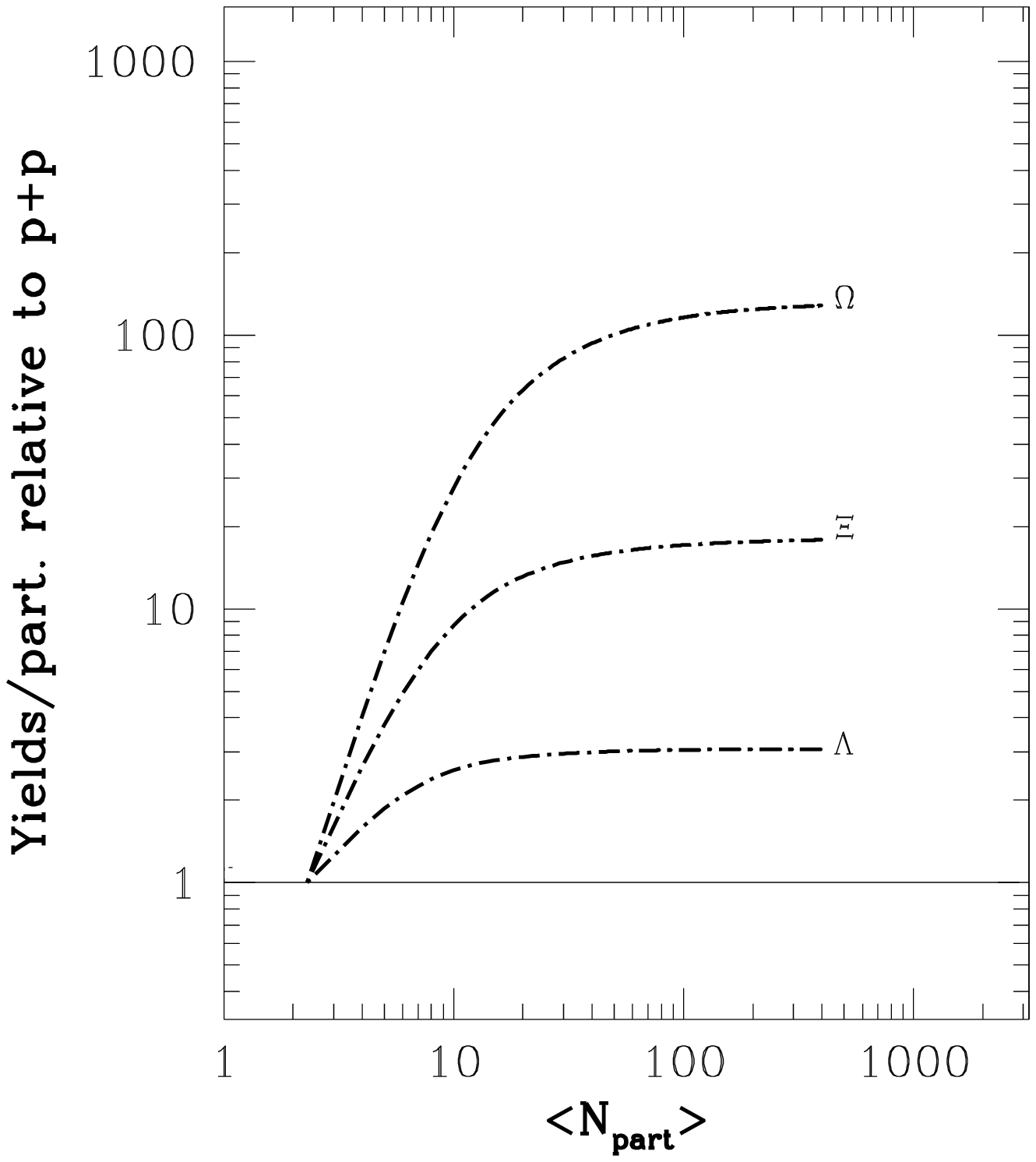}
\vskip -0.4cm 
\caption{Centrality dependence of the relative
enhancement
 of particle yields/participant in central Pb-Pb to p-p collisions
at fixed energy $\sqrt{s}=8.73$ GeV. \hfill}\label{oxl}
\end{minipage}
\hspace{\fill}
\begin{minipage}[t]{74mm}
\epsfig{width=7.4cm,figure=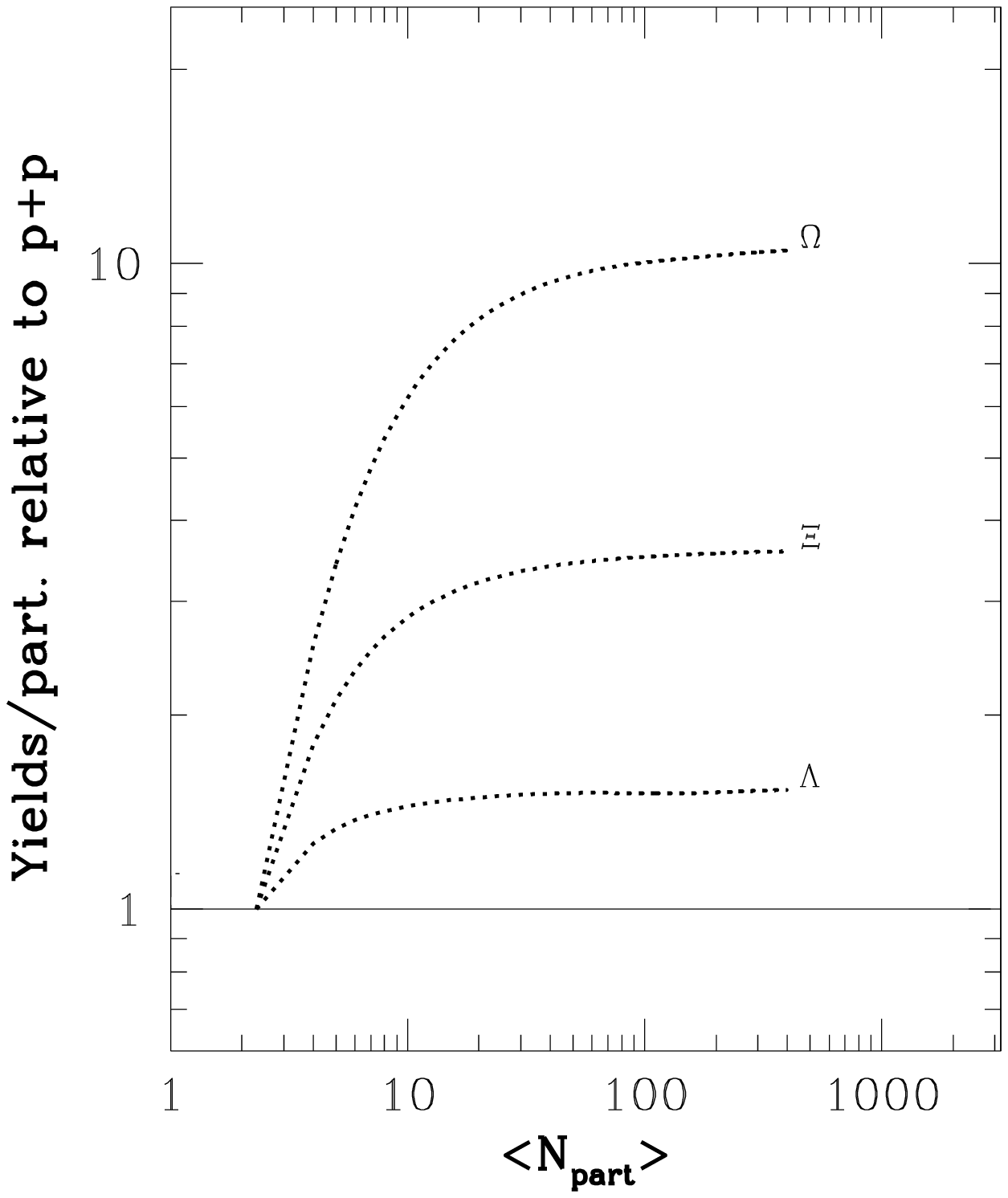}
\vskip -0.4cm 
\caption{ Centrality dependence of the relative
enhancement of particle yields/participant in central Pb-Pb to p-p
collisions at fixed energy $\sqrt{s}=130$ GeV. \hfill}
\label{oxlr}
\end{minipage}
\end{figure}
\begin{figure}[t]
\vskip -1.cm 
\begin{minipage}[t]{80mm}
\epsfig{width=7.5cm,figure=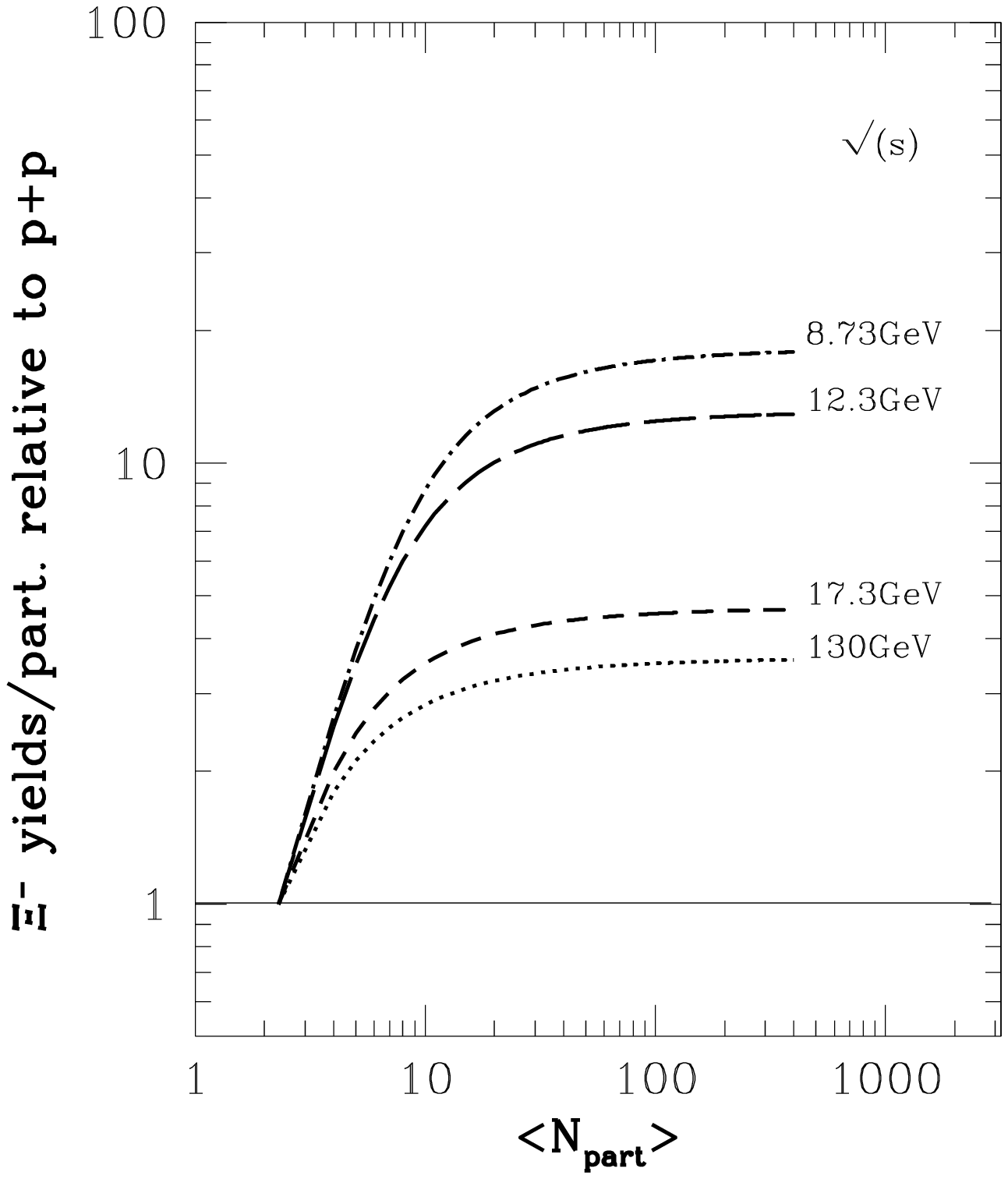}
\vskip -0.4cm \caption{ Centrality dependence of relative
enhancement of $\Xi^-$ yields/participant in central Pb-Pb  to p-p
reactions at different collision energies. \hfill}\label{xi}
\end{minipage}
\hspace{\fill}
\begin{minipage}[t]{76mm}
\epsfig{width=7.5cm,figure=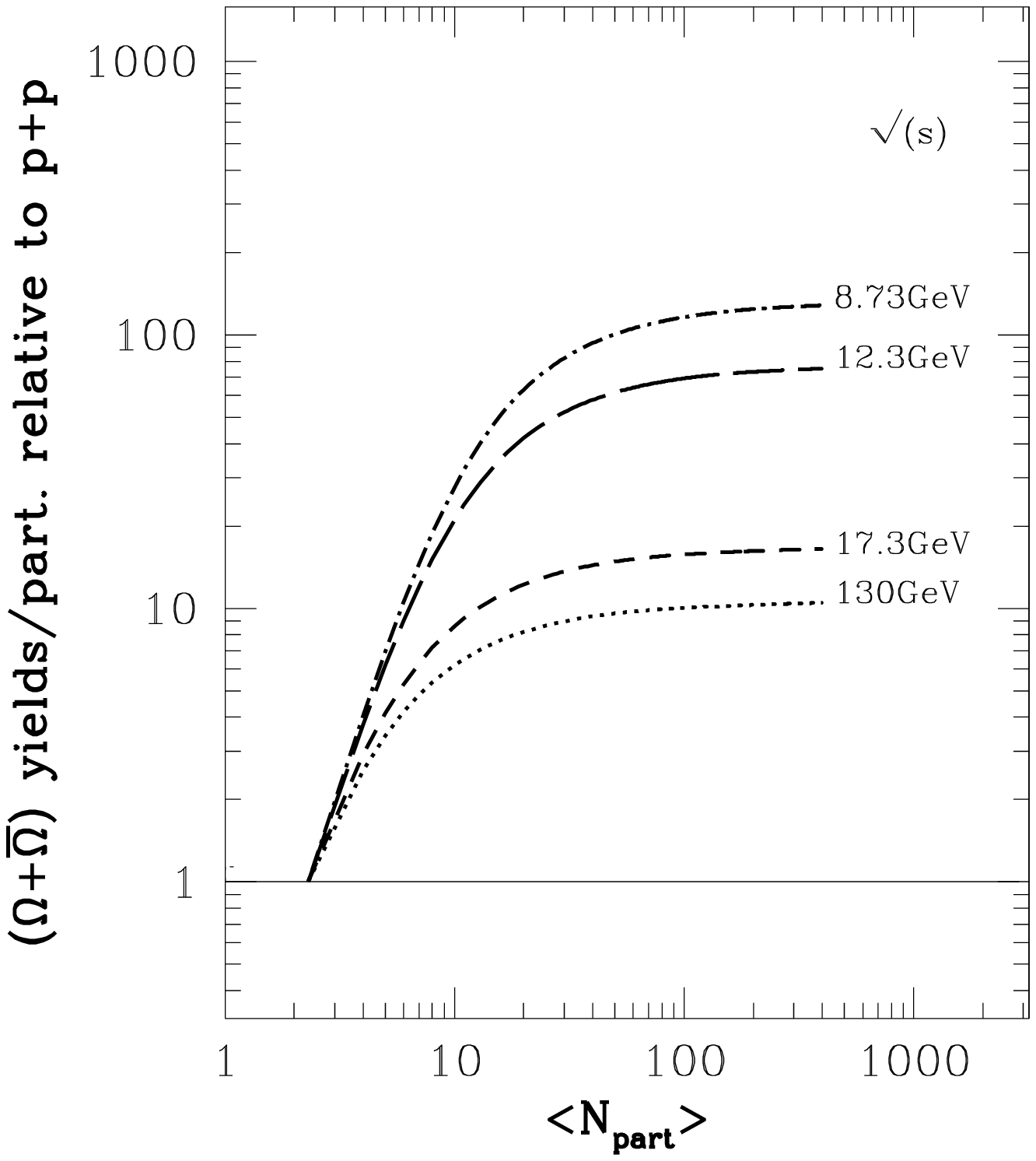}
\vskip -0.4cm 
\caption{Centrality dependence of relative
enhancement of $({\Omega}+\bar{\Omega})$ yields/participant in
central Pb-Pb to p-p reactions at different collision energies.
\hfill} \label{omega}
\end{minipage}
\end{figure}

In Fig.\,\,\ref{oxl} and Fig.\,\,\ref{oxlr}
 we show the results on relative (multi)strange
baryon enhancement from p-p to Pb-Pb collisions  at $\sqrt s=8.73$
GeV  and
at $\sqrt s=130$ GeV respectively. It is clear that  the same
enhancement pattern as at SPS is expected in SM  to
appear for all relevant energies. To see the  dependence of the
strength of the enhancement with energy we show in Fig.\,\,\ref{xi}
and Fig.\,\,\ref{omega} the relative enhancement of $\Xi$ and
$(\Omega+\bar\Omega )$   for different collision energies. The
enhancement is the largest at lowest energy,  for $\Omega$ it can
be even by a factor of almost ten larger at  $\sqrt s=8.73$ than observed
at SPS top energy. At RHIC  enhancement of $\Omega$ is smaller than at
 SPS. These predictions are in contrast with UrQMD\cite{bleicher}
which predicts an enhancement  at RHIC larger by a factor of four      than at SPS. Note that at all energies all figures display the WA97 pattern: hierarchy
and saturation, the latter indicating that the grand canonical limit is reached.
This pattern was predicted as a signal for quark-gluon plasma
formation \cite{rafelski1}. In the context of the  considered SM 
the enhancement pattern of (multi)strange baryons should be
observed at all SPS  energies,  with increasing strength toward
lower beam energy. Thus, the results of the above SM makes it
clear  that strangeness enhancement and enhancement pattern is not
a unique
signal of deconfinement as these  features are expected
to be also there at energies  where QGP is not expected to be formed.

The quantitative results shown in Fig.\,\,\ref{oxl} to Fig.\,\,\ref{omega}
contain some uncertainties. The magnitude of the enhancement is
very sensitive to temperature taken at given collision energy.
Changing $T$ by 5 MeV, a typical error on $T$ in thermal analysis,
can change, e.g., the enhancement of $\Omega$ shown in Fig.\,\,\ref{oxl}
by a factor of two. The $N_{part}$ dependence of strange baryon enhancement seen in
Figs. 3-6 was obtained assuming linear dependence of volume
parameter $V$ in Eq.\,(\ref{vol}) with $N_{part}$. In general
 $V$ could have a
weaker  dependence with centrality which could be reflected in
only moderate  increase of the enhancement with centrality and
saturation appearing at  larger volume. In addition, including
variation of thermal parameters, in particular the baryon-chemical
potential, with centrality or extending the model to canonical
description of baryon number conservation \cite{b1} or, finally,
including a possible asymmetry between strangeness
under-saturation factor in pp and AA collisions \cite{b1,bc},
could change our numerical values.
However, independently of these uncertainties, the main results:
(i) enhancement decreasing with increasing collision energy and
(ii) enhancement pattern being  preserved at all SPS and RHIC energies, are always valid.

\section{Concluding remarks}
 We have  shown that in terms of statistical model
the  relative enhancement of (multi)strange baryons from
proton-proton to nucleus-nucleus collisions is a decreasing
function of collision energy. Experimentally this fact was already
obtained for  kaon yields and is shown to be   expected for
multistrange baryons.  In addition, an increase of the enhancement
with strangeness content of the particle is a  generic feature of
our model,  independent of collision energy. On the qualitative
level  the only input being required in the model to make the
above predictions is the information that freezeout temperature is
increasing with collision  energy and that the chemical potential
shows the opposite dependence. The above required conditions are
well confirmed by a very detailed analysis of particle production
at different collision energies.

We have presented the quantitative predictions for relative
enhancement of  $\Lambda$, $\Xi$  and $\Omega$ yields in the
energy range form $\sqrt s=8.7$ up to $\sqrt s=130$ GeV. We have
discussed possible uncertainties of presented  results. The
relative enhancement at RHIC was found to be lower than at the
SPS. This is in contrast with UrQMD \cite{bleicher}
predictions and with the  previous qualitative predictions of
observed enhancement as being entirely due to quark-gluon plasma
formation \cite{rafelski1}. Even an abrupt change 
in the enhancement of $\bar\Xi/N_{part}$ versus $N_{part}$,  recently reported by the NA57, could be possibly accounted for in
terms of canonical  model when assuming a very particular centrality
dependence of the correlation volume \cite{muller}.
 It would correspond to a sudden
jump of the volume, as in a first order phase transition,
which can be taken into account in the canonical suppression factor. 

\section*{Acknowledgments}
Interesting discussions with B. M\"uller and   H. Satz are
acknowledged. (K.R.) also acknowledges a partial support of  the
Polish Committee for Scientific Research (KBN-2P03B 03018). 
(A.T.) also wishes to thank M. Gonin and the organizing committee
of CIPPQG for a good atmosphere at Ecole Polytechnique
and for an excellent Conference dinner at the Palais du Luxembourg.

\end{document}